\documentclass[12pt,hyperref,notitlepage,amsfont,showpacs,showkeys]{revtex4-1}

\usepackage[]{graphicx}
\graphicspath{{figures/}}
\usepackage{amsmath,amssymb}

\newcommand{\E}{E_{0}}
\newcommand{\rhos}{\rho_{s,600}}

\newcommand{\degr}{^{\circ}}
\newcommand{\depthunit}{g/cm$^2$}
\newcommand{\xmax}{x_{\text{max}}}
\newcommand{\obslev}{_{\text{obs.}}}
\newcommand{\ionize}{_{\text{ion.}}}
\newcommand{\Eel}{E_{\text{el.}}}

\newcommand{\QGS}{QGSJet01d}
\newcommand{\QGSII}{QGSJet-II-04}
\newcommand{\SIBYLL}{SIBYLL-2.1}
\newcommand{\EPOS}{EPOS-LHC}

\begin{document}

\title{Revision of the energy calibration of the Yakutsk EAS array}

\author{A. V. Glushkov}

\author{M. I. Pravdin}

\author{A. Sabourov}
\email{tema@ikfia.sbras.ru}

\affiliation{Yu. G. Shafer Institute of cosmophysical research and aeronomy}
\address{677980, Lenin Ave. 31, Yakutsk, Russia}

\begin{abstract}
    Responses of surface and underground scintillation detectors of Yakutsk array are calculated for showers initiated by primary particles with energy $\E \ge 10^{17}$~eV within the frameworks of \QGS, \QGSII, \SIBYLL{} and \EPOS{} hadron interaction models. A new estimation of $\E$ is obtained with the use of various methods. The resulting energy is lower compared to the obtained with earlier method by factor $\sim 1.33$.
\end{abstract}



\maketitle

\section{Introduction}

    Energy spectrum of ultra-high energy cosmic rays (UHECR), cosmic rays (CR) with $\E \ge 10^{17}$~eV energy, is one of the main links in the chain of complex problems associated with understanding the nature of primary particles with such energy. Mechanisms of their production and acceleration in astrophysical sources and various effects they experience during the propagation in the Universe have direct effect on the observed primary CR spectrum. Recently, a significant progress has been achieved in interpretation of its structural features in the ultra-high energy domain. The black-body cutoff at $\E \sim 6 \times 10^{19}$~eV predicted by Greisen~\cite{Greisen:PRL(1966)} and Zatsepin and Kuz'min~\cite{Zatsepin&Kuzmin:LETPLett(1966)} (the GZK cutoff) has been confirmed~\cite{GZK-HiRes:PRL(2008), GZK-PAO:PRL(2008)}, thus pointing at extragalactic origin of the most energetic CR particles. The second knee (at $\sim 10^{17}$~eV) and the ankle (at $\sim 10^{18}$~eV) are commonly associated with a transition between galactic and extragalactic CR components; and though its exact location on the energy scale is not known precisely, there are plenty of theoretical scenarios compatible with existing experimental observations (e.g.~\cite{Aloisio-etal:AstropartPhys(2007), Berezhko-etal:Astropart.Phys.(2012), Zirakashvili&Ptiskin:Bull.Russ.Acad.Sci.:Phys(2009)}).

However, there is certain discrepancy in the world array of experimental results. CR spectra measured by various UHECR experiments~\cite{Edge_etal:J.Phis.A(1973), Yakutsk:Proc20thICRC(1987), AGASA:Proc27thICRC(2001), HiRes:Astropart.Phys.(2005), PAO:PhysLettB(2010)} confirm such spectral features as the ankle and the second knee, but they differ from each other in absolute intensity by factor of almost $2$~\cite{Glushkov&Pravdin:JETPLett(2008), Int.CRS.Workgroup:EPJ(2013)}. In particular, the spectrum measured by Yakutsk experiment lies above all the world data. In this context, the data published by the Yakutsk group signify the upper limit of the spectrum intensity, and data from the Pierre Auger Observatory (PAO)~--- its lower limit.

Such situation to a large extent stems from the fact that the only available method of UHECR observation is indirect, conducted by registering cascades of secondary particles produced by primary UHECRs in the Earth's atmosphere: extensive air showers (EAS). Most of largest experiments employ differing observational techniques and, consequently, rely on different methods to reconstruct the energy of primary particles. Hence, one cannot do without theoretical notion of EAS development.

The Yakutsk EAS array stands out from other large arrays for its complexity: since it is equipped with detectors of three types, it simultaneously registers several shower components. Charged particles (electrons, positrons and muons) are recorded with 2~m$^2$ surface scintillation detectors (SSD). Muon component arising from nuclear interactions is registered with detectors of the same type placed below the ground level, in order to prevent electromagnetic contamination by creating a shield with $1\times \sec{\theta}$~GeV threshold. Cherenkov light emitted by EAS charged particles is recorded with integral Cherenkov detectors based on the FEU-49 photomultiplier tube.

Cherenkov component carries information about $\sim 80$\,\% of primary energy dissipated in the atmosphere and, thus, enables to determine $\E$ with calorimetric method~\cite{Yakutsk:VI-ECRS(1978), Glushkov:PhD(1982), Yakutsk:Bull.Acad.Sci.USSR(1991), Yakutsk:TokyoWorkshop(1993), Yakutsk:Proc28thICRC(2003)}. This method defines the $\E$ as a sum of energies of all EAS components and connects it with experimentally measured value $\rhos$ (it will be discussed in greater detail in Section~\ref{S3}). Originally, it was introduced in~\cite{Yakutsk:ProcLaPaz(1962)} for energies $\sim 10^{15}$~eV. In Yakutsk experiment it was applied to showers with $\E \simeq (1.0 - 100) \times 10^{17}$~eV at zenith angles $\theta \le 45\degr$~\cite{Yakutsk:VI-ECRS(1978), Glushkov:PhD(1982)} and resulted in the following approximation for primary energy reconstruction:
\begin{eqnarray}
    \E = & (4.1 \pm 1.4) \times 10^{17} \cdot (\rhos(0\degr))^{0.97 \pm 0.04}\text{~(eV),}
    \label{eq:EnergyEst1982} \\
    &\rhos(0\degr) = \rhos(\theta) \times \exp\left(\frac{(\sec{\theta} - 1) \cdot x_0}{\lambda_{\rho}}\right)\text{,}
    \label{eq:rho_vert1982} \\
    &\lambda_{\rho} = 400 \pm 45 \text{\quad (\depthunit),}
    \label{eq:FreePath1982}
\end{eqnarray}
where $x_0 = 1020$~\depthunit{}, $\rhos(\theta)$ is the density of charged particles (m$^{-2}$) measured by SSDs at the distance $R = 600$~m from shower axis and $\lambda_{\rho}$ is attenuation length. Later, the relations~(\ref{eq:EnergyEst1982}) and (\ref{eq:FreePath1982}) were changed slightly (see~\cite{Yakutsk:Bull.Acad.Sci.USSR(1991), Yakutsk:TokyoWorkshop(1993), Yakutsk:Proc28thICRC(2003)}):
\begin{eqnarray}
    \E & = & (4.8 \pm 1.6) \times 10^{17} \cdot (\rhos(0\degr))^{1.0 \pm 0.02}\text{,}
    \label{eq:EnergyEst1993}\\
    \lambda_{\rho} & = & (450 \pm 44) + (32 \pm 15) \cdot \log_{10}{\rhos(0\degr)}\text{.}
    \label{eq:FreePath1993}
\end{eqnarray}
The intensity of the CR energy spectrum estimated with the use of~(\ref{eq:EnergyEst1993}) turned out to be significantly higher than the world data (see e.g.~\cite{Tunka:Astropart.Phys.(2013)}). In~\cite{Dedenko_etal:(2007)} estimation of $\E$ for Yakutsk data was presented obtained for primary protons within the framework of QGSJet01 model, which was 1.6 times lower than (\ref{eq:EnergyEst1993}). Here we consider energy calibration of registered showers based on modern CORSIKA code (version 6.7370)~\cite{CORSIKA}.

\section{Lateral distribution of the detector signal}

Basic EAS parameters measured in Yakutsk experiment (arrival direction, coordinates of the shower axis, primary energy) are reconstructed with lateral distribution function (LDF) of all particles (electrons, muons and high-energy photons) which are registered by SSDs. These particles pass through a multilayer shield consisting of snow, iron, wood and aluminium (total thickness amounts to $\sim 2.5$~\depthunit) and then~--- through a $5$~cm thick scintillator (with the density $\sim 1.06$~g/cm$^3$). The energy deposit in a scintillator $\Delta E_{s}(R)$ is proportional to the number of particles passed though a detector and is measured in relative units:
\begin{equation}
    \rho_s(R) = \frac{\Delta E_{s}(R)}{E_1}\text{\quad(m$^{-2}$),}
    \label{eq:RelUnit}
\end{equation}
where $E_1 = 11.75$~MeV, which is the energy released in a scintillator during the passage of a vertical relativistic muon (the response unit).

Scintillation detectors are calibrated and controlled with the use of amplitude density spectra from background cosmic particles~\cite{Yakutsk:Ya.Fil.SB.Acad.Sci.USSR(1974)}. Herewith, the integral spectra of two types are used. First one~--- a spectrum from a single detector, which is controlled by a nearby detector mounted in the same station (the so-called ``spectrum of double coincidence'' with the frequency $\simeq (2-3)$~s$^{-1}$). Second one~--- uncontrolled spectrum with the frequency $\sim 200$~s$^{-1}$, which is used to calibrate muon detectors. Both spectra are described by a power law:
\begin{equation}
    F(>\rho) \sim \rho^{-\eta} \sim U^{-\eta}\text{,}
    \label{eq:CalibrationSpectra}
\end{equation}
where values of $\eta$ for both spectra were obtained experimentally. For spectrum of the first type $\eta = 1.7$ and for spectrum of the second type $\eta = 3.1$. $\rho = U / U_1$~--- particle density measured in units of signal amplitude $U_1$ from a reference detector during the passage of a vertical relativistic cosmic muon. The procedure of calibration and control consists of continual monitoring of the $U_1$ value in all detectors by periodical measurements of their density spectra. The procedure is performed once a two days. Spectra of double coincidence are collected for $2$~hours, uncontrolled spectra~--- for 30~minutes.

Within the framework of models \QGS~\cite{QGSJet01}, \QGSII~\cite{QGSJetII}, \SIBYLL~\cite{SIBYLL} and \EPOS~\cite{EPOS} we calculated LDF of the SDD response in showers, initiated by primary protons and iron nuclei with energies $10^{17.0} - 10^{19.5}$~eV arriving at different zenith angles. FLUKA package~\cite{FLUKA:Proc.AIP(2007), FLUKA:Man} was chosen for treatment of lower energy interactions. At first, the response $u_m(\epsilon,\theta)$ from a single particle of a type $m$ (where $m$ is electron, muon or gamma-photon) with energy $\epsilon$ was calculated. During the calculation, all the main precesses occurring in the detector during energy release/consumption with corresponding cross-sections were put into consideration: ionization and bremsstrahlung~-- for charged particles; pair production and delta-electrons from Compton effect~-- for gamma-photons. Then the development of air shower was simulated with CORSIKA code. For each set of primary parameters (mass of primary particle, its energy and incident zenith angle) 500 showers were simulated. In order to speed-up the simulation, the ``thin-sampling'' mechanism, introduced in~\cite{EGS4}, was activated in the CORSIKA code~\cite{CORSIKA:UG,CORSIKA}. The thinning level $\epsilon_{\text{th.}} = \epsilon_{\text{min}} / E_0$, controlling the minimal energy of secondary particles $\epsilon_{\text{min}}$ treated by CORSIKA, was defined in the interval $3.16 \cdot 10^{-6}, 10^{-5}$ and weight limit of secondary particles $w_{\text{max}}$~--- in the interval $10^4, 3.16 \cdot 10^{6}$, depending on the primary energy. This was done in order to limit the growth of artificial fluctuations induced by thin-sampling in showers with lower energies.

During conversion to density, the number of particles was calculated in the detector of a given area. Resulting showers were averaged together and mean energy spectra $d_m(\epsilon, R, \theta)$ were calculated for all particle types in intervals $(\log_{10}{R_{j}}, \log_{10}{R_{j} + 0.04})$. The signal~(\ref{eq:RelUnit}) at a distance $R$ was defined as a sum of responses:
\begin{equation}
    \rho_s(R) = \sum_{m}^{3} \sum_{i = 1}^{I_{m}} u_m(\epsilon_i,\theta_i) d_{m}(\epsilon_i, R, \theta_i)\text{,}
    \label{eq:SurfaceSignal}
\end{equation}
where $I_m$~--- the number of particles of a type $m$ hitting a detector at a distance $R$.

\begin{figure}
    \centering
    \includegraphics[width=0.85\textwidth, clip]{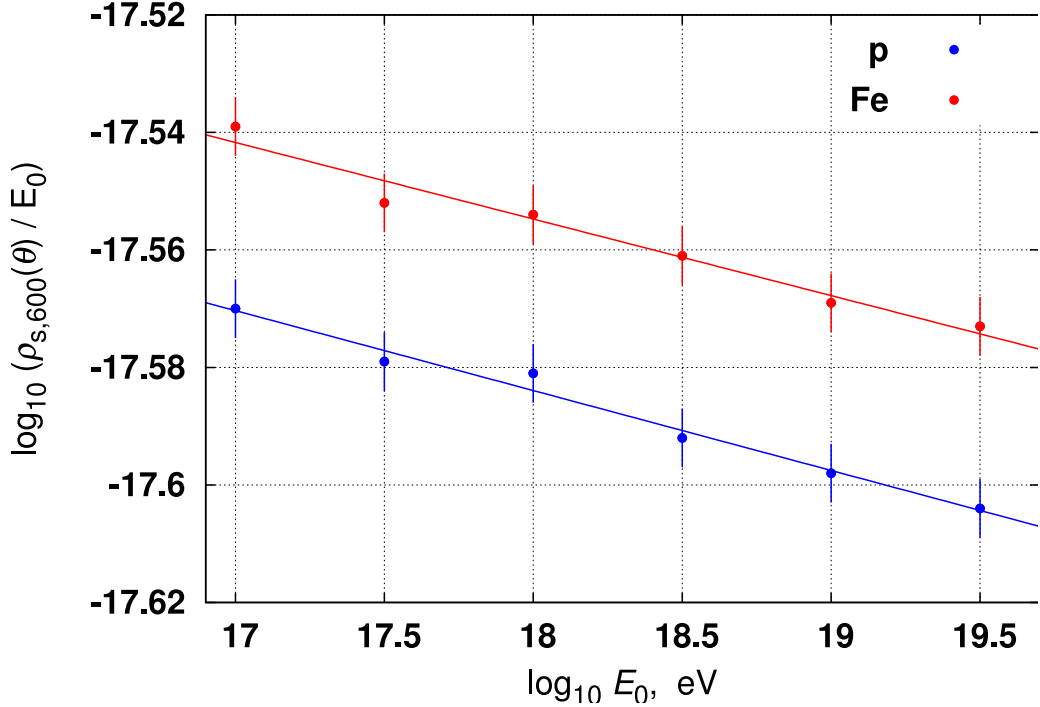}
    \caption{Energy dependence of $\log_{10} \left({\rhos(\theta)/\E}\right)$ for primary protons and iron nuclei according to predictions of QGSJet01d for vertical showers.}
    \label{fig:s600_mc}
\end{figure}

On Fig.\ref{fig:s600_mc} the dependence of the value $\log_{10} \left(\rhos(0\degr) / \E\right)$ from $\E$ is shown for primary protons (open circles) and iron nuclei (black circles) as predicted by \QGS{} model. They satisfy the relation:
\begin{equation}
    \E = (3.24 \pm 0.1) \times 10^{17} \cdot (\rhos(0\degr))^{1.015}\text{,}
    \label{eq:EnergyQGS01}
\end{equation}

Other models~--- \QGSII, \SIBYLL{} and \EPOS~--- give the following estimations correspondingly:
\begin{eqnarray}
    \E & = & (3.52 \pm 0.1) \times 10^{17} \cdot (\rhos(0\degr))^{1.02}\text{,}
    \label{eq:EnergyQGSII} \\
    \E & = & (3.09 \pm 0.1) \times 10^{17} \cdot (\rhos(0\degr))^{1.015}\text{,}
    \label{eq:EnergySIBYLL} \\
    \E & = & (3.74 \pm 0.1) \times 10^{17} \cdot (\rhos(0\degr))^{1.02}\text{,}
    \label{eq:EnergyEPOS}
\end{eqnarray}

Averaging over all models gives the dependence:
\begin{equation}
    \E = (3.40 \pm 0.18) \time 10^{17} \cdot (\rhos(0\degr))^{1.017}\text{,}
    \label{eq:EnergySSD}
\end{equation}
which resulted in a lower estimated value of $\E$ by factor $1.20$ when compared to (\ref{eq:EnergyEst1982}) and by $1.41$ when compared to (\ref{eq:EnergyEst1993}).

\begin{figure}
    \centering
    \includegraphics[width=0.85\textwidth, clip]{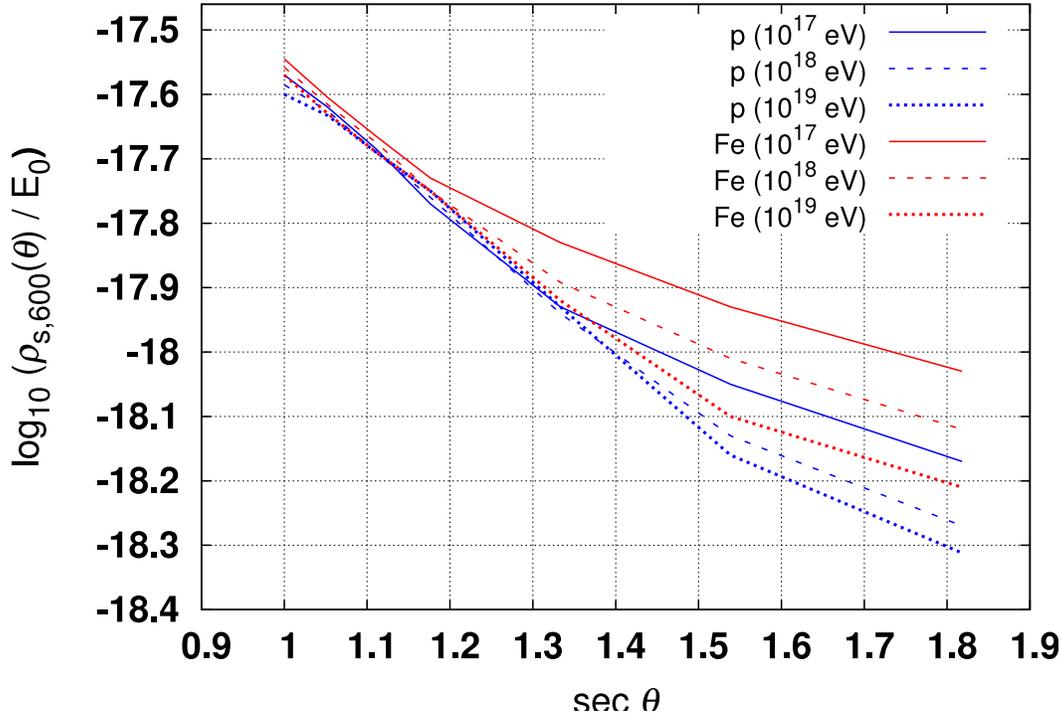}
    \caption{Zenith-angular dependence of $\log_{10} \left({\rhos(\theta)/\E}\right)$ according to QGSJet01d predictions for protons and iron nuclei with energy $\E=10^{17}, 10^{18}$ and $10^{19}$~eV.}
    \label{fig:s600_angular}
\end{figure}

Zenith-angular dependences of $\log_{10} \left({\rhos(\theta)/\E}\right)$ according to \QGS{} model are shown on Fig.\ref{fig:s600_angular}. They satisfy a linear dependency with $\lambda_{\rho} = 415 \pm 15$~\depthunit{} at any composition of primary CR when $\sec{\theta}$ is lesser than:
\begin{equation}
    \sec{\theta}_{\text{lim.}} = a + b \log{\rhos(\theta)}\text{,}
    \label{eq:ZenithLim}
\end{equation}
where $a = 1.26$ and $b = 0.077$. In the case of protons the relation (\ref{eq:ZenithLim}) is agreeable for inclined showers with $a = 1.477$, at $\E = 10^{19}$~eV, with attenuation length $\lambda_{\rho} = 415$~\depthunit{} and for $\theta \le 50\degr$. In other cases the dependency is more complex.

\section{Calorimetric method}\label{S3}

\begin{table*}
    \caption{Observables of EAS with $\E = 10^{18}$~eV and $\cos{\theta} = 0.95$ from primary nuclei ($A$) according to CORSIKA~\cite{CORSIKA} simulation and experiment~\cite{Glushkov:PhD(1982)}.}
    \label{t:1}
    \centering
    \begin{tabular}{lccccccc}
        \hline
        \hline
        & & $k_{\gamma}(\theta)$ & $k\ionize(\theta)$ & $F(\theta)$ & $\left<N_s(\theta)\right>$ &
            $\rhos(\theta)$ & $\left<N_{\mu}(\theta)\right>$ \\
        model & $A$ & $(\times 10^4)$ & $(\times 10^4)$ & $(\times 10^{13})$ & $(\times 10^8)$ &
            & $(\times 10^6)$ \\
        & & eV$^2$ & eV$^2$ & eV$^{-1}$ & & m$^{-2}$ & \\
        \hline
        QGSJet01d              & p    & $0.341$ & $2.846$ & $2.104$ & $2.178$ & $2.312$ & $5.000$ \\
                               & Fe   & $0.224$ & $2.910$ & $2.148$ & $1.250$ & $2.432$ & $7.225$ \\
        \hline
        QGSJet-II-04           & p    & $0.364$ & $2.816$ & $2.070$ & $2.296$ & $2.438$ & $5.582$ \\
                               & Fe   & $0.246$ & $2.894$ & $2.148$ & $1.358$ & $2.636$ & $7.777$ \\
        \hline
        SIBYLL-2.1             & p    & $0.345$ & $2.822$ & $2.100$ & $2.512$ & $2.193$ & $4.254$ \\
                               & Fe   & $0.224$ & $2.910$ & $2.228$ & $1.384$ & $2.249$ & $4.930$ \\
        \hline
        EPOS-LHC               & p    & $0.377$ & $2.815$ & $2.023$ & $2.355$ & $2.655$ & $5.905$ \\
                               & Fe   & $0.230$ & $2.894$ & $2.133$ & $1.419$ & $2.917$ & $8.180$ \\
        \hline
                               & p    & $0.357$ & $2.825$ & $2.074$ & $2.335$ & $2.400$ & $5.185$ \\
        average                & Fe   & $0.231$ & $2.902$ & $2.164$ & $1.353$ & $2.558$ & $7.028$ \\
                               & p-Fe & $0.294$ & $2.864$ & $2.119$ & $1.844$ & $2.479$ & $6.107$ \\
        \hline
        experiment~\cite{Glushkov:PhD(1982)} & --   & \multicolumn{2}{c}{$3.700$}
                                                          & $2.510$ & $1.793$ & $2.656$ & $6.000$ \\
        \hline
        \hline
    \end{tabular}
\end{table*}

We considered the energy balance starting from the example of experimental data from~\cite{Yakutsk:VI-ECRS(1978), Glushkov:PhD(1982)}. Earlier, these data had provided a basis for the calorimetric method of $\E$ estimation adopted for the Yakutsk array. The observables and main components constituting the primary energy are given in Tables~\ref{t:1} and \ref{t:2} for $\E = 10^{18}$~eV and $\cos{\theta} = 0.95$. The $F$ column in the Table~\ref{t:1} is the flux of Cherenkov photons measured with integral Cherenkov light detectors. The values for $k_{\gamma}$ and $k\ionize$ in the same table were obtained in simulation with CORSIKA. Mean values of $N_{s}$ and $N_{\mu}$ were obtained from the LDFs averaged over energy interval. The row entitled ``average p-Fe''corresponds to values averaged over all models and compositions. The energy dissipated in the atmosphere by electromagnetic component equals to
\begin{equation}
    E_i = E_{\gamma} + E\ionize\text{,}
    \label{eq:EnergyEM}
\end{equation}
where $E_{\gamma}$ is energy of gamma-photons on observation level, $E\ionize$~--- summary ionization losses of all electrons and positrons. It is proportional to the total flux $F$ of Cherenkov radiation in the atmosphere:
\begin{equation}
    E_i = k \cdot F\text{,}
    \label{eq:CherenkovEM}
\end{equation}
where $k$~(eV/photon eV$^{-1}$) is the scaling factor:
\begin{equation}
    k = k_{\gamma} + k\ionize = \frac{E_{\gamma} + E\ionize}{F}\text{.}
    \label{eq:scaling}
\end{equation}

\begin{table*}
    \caption{Energy balance of EAS with $\E = 10^{18}$~eV and $\cos{\theta} = 0.95$ from primary $(A)$ according to CORSIKA~\cite{CORSIKA} simulation and experiment~\cite{Glushkov:PhD(1982)}.}
    \label{t:2}
    \centering
    \begin{tabular}{lccccccc}
        \hline
        \hline
        & & $E_{\gamma}$ & $E\ionize$ & $E_{el}$ & $E_{\mu}$ & $\Delta E$ & $\E$ \\
        model & $A$ & $(\times 10^{17})$ & $(\times 10^{17})$ & $(\times 10^{17})$ & $(\times 10^{17})$ &
        $(\times 10^{17})$ & $(\times 10^{17})$ \\
        & & eV & eV & eV & eV & eV & eV \\
        \hline
        QGSJet01d   & p    & $0.806$ & $6.620$ & $1.469$ & $0.517$ & $0.565$ & $9.978$ \\
                    & Fe   & $0.529$ & $6.660$ & $1.306$ & $0.785$ & $0.798$ & $9.972$ \\
        \hline
        QGSJetII-04 & p    & $0.859$ & $6.476$ & $1.474$ & $0.547$ & $0.624$ & $9.980$ \\
                    & Fe   & $0.582$ & $6.430$ & $1.302$ & $0.844$ & $0.866$ & $9.981$ \\
        \hline
        SIBYLL-2.1  & p    & $0.909$ & $6.625$ & $1.523$ & $0.428$ & $0.491$ & $9.976$ \\
                    & Fe   & $0.528$ & $6.679$ & $1.340$ & $0.702$ & $0.716$ & $9.965$ \\
        \hline
        EPOS-LHC    & p    & $0.891$ & $6.412$ & $1.482$ & $0.524$ & $0.657$ & $9.966$ \\
                    & Fe   & $0.543$ & $6.415$ & $1.305$ & $0.794$ & $0.898$ & $9.955$ \\
        \hline
        average     & p    & $0.866$ & $6.533$ & $1.487$ & $0.504$ & $0.584$ & $9.974$ \\
                    & Fe   & $0.546$ & $6.531$ & $1.313$ & $0.781$ & $0.820$ & $9.968$ \\
                    & p-Fe & $0.706$ & $6.532$ & $1.400$ & $0.643$ & $0.702$ & $9.970$ \\
        \hline
        experiment~\cite{Glushkov:PhD(1982)} 
        & --   & \multicolumn{2}{c}{$9.287$}
                                               & $0.947$ & $0.636$ & $0.860$ & $11.730$ \\
        \hline
        new estimation & -- & \multicolumn{2}{c}{$7.926$}
                                               & $0.947$ & $0.618$ & $0.702$ & $10.190$ \\
        \hline
        \hline
    \end{tabular}
\end{table*}

\begin{figure}
    \centering
    \includegraphics[width=0.85\textwidth, clip]{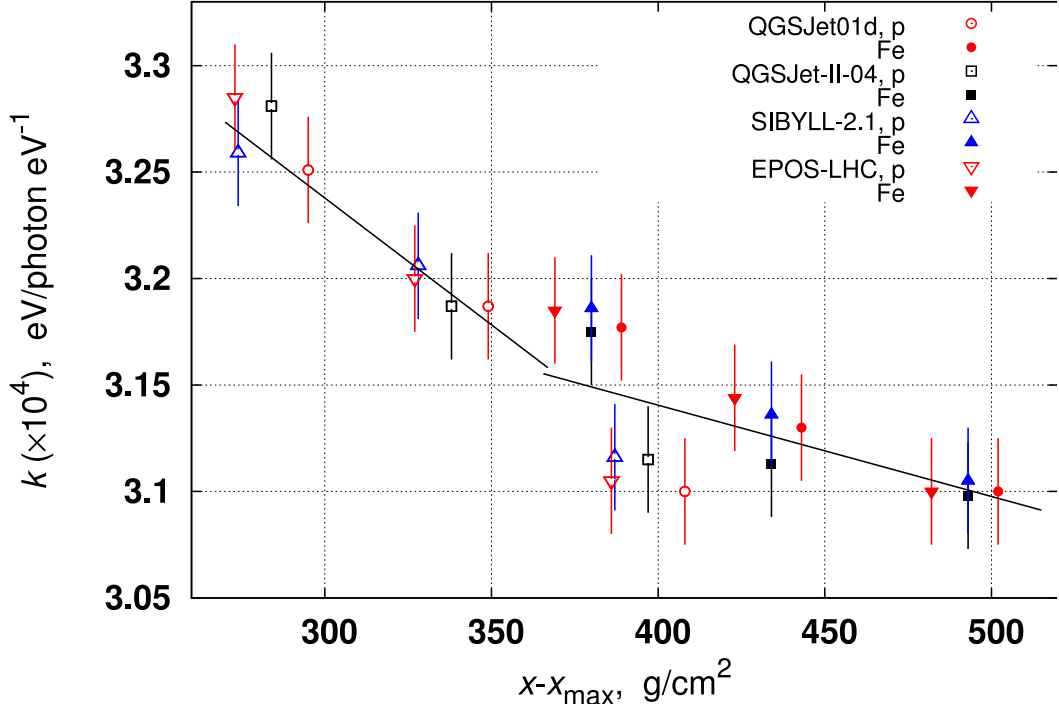}
    \caption{Dependence of the scaling ratio~(\ref{eq:scaling}) from the path between $\xmax$ and $x\obslev$ for two CR compositions. Lines represent approximations.}
    \label{fig:scaling}
\end{figure}

On Fig.\ref{fig:scaling}, the dependence of the scaling factor~(\ref{eq:scaling}) from the path from $\xmax$ to observation level ($x\obslev = x_0 \times \sec{\theta}$~\depthunit) is shown. The flux $F$ is determined with respect to its attenuation by factor $1.15$ due to Rayleigh scattering in clean atmosphere and degradation of the relative transparency in sampling events~\cite{Yakutsk:VI-ECRS(1978), Glushkov:PhD(1982)} by factor $1.1$. It is given for radiation interval $1$~eV:
\begin{equation}
    F = \frac{1.265 F\obslev}{\Delta\epsilon}\text{,}
    \label{eq:CherenkovFlux}
\end{equation}
where $F\obslev$ is the flux measured in experiment with integral Cherenkov light detectors and
\begin{equation}
    \Delta\epsilon = 12400 \cdot \left(\frac{1}{\lambda_1} - \frac{1}{\lambda_2}\right) \simeq 2.58\text{\quad (eV).}
    \label{eq:CherenkovBand}
\end{equation}
Here $\lambda_1 = 3000$~\AA, $\lambda_2 = 8000$~\AA. The energy $\Eel$ is the amount of primary energy carried by electrons and positrons to the observation level. It was estimated by integrating the differential energy deposit over the cascade curve $N_e(x)$ below the observation level $x\obslev$:
\begin{align}
    \Eel & = \int_{x\obslev}^{\infty} \left(\frac{\text{d}E}{\text{d}x}\right)\ionize \cdot N_e(x) \text{d}x
         \simeq \nonumber \\
         &\simeq 2.2 \times 10^{6} \cdot N_e(x\obslev) \times \nonumber \\
         &\qquad{} \int_{x\obslev}^{\infty} \exp\left(\frac{x\obslev - x}{\left<\lambda_N\right>} \right) \text{d}x\text{,}
    \label{eq:EnergyTransf}
\end{align}
where $\left<\lambda_N\right> \simeq 240$~\depthunit{}. $N_e(x\obslev)$ is the number of electrons at observational level, which was determined from the relation:
\begin{equation}
    N_e(x\obslev) \simeq \left<N_s(x\obslev)\right> - 1.8 \cdot \left<N_{\mu}(x\obslev)\right>\text{,}
    \label{eq:NeEstimation}
\end{equation}
where $\left<N_s(x\obslev)\right>$ and $\left<N_{\mu}(x\obslev)\right>$ are mean values of the total number of responses from all particles and muons with $1$~GeV threshold, obtained by integrating of experimentally measured corresponding LDFs~\cite{Yakutsk:VI-ECRS(1978),Glushkov:PhD(1982)}. The ratio $1.8$ accounts the difference between the numbers of muons measured by SSDs and underground detectors with $1$~GeV threshold. It was derived from earlier calculations~\cite{Glushkov:PhD(1982)} and is roughly agreeable with present simulation.

Energy of muons was measured experimentally:
\begin{equation}
    E_{\mu} = \left<E_{1\mu}\right> \cdot \left<N_{\mu}(x\obslev)\right>\text{,}
    \label{eq:MuonEnergy}
\end{equation}
where $\left<E_{1\mu}\right> = 10.6$~GeV, which is the mean energy of a single muon.

From the data given in Table~\ref{t:2}, averaged over all models, the summary value $E_i + \Eel + E_{\mu}$ amounts to $\simeq 93$\,\% from primary energy. The rest of it ($\Delta E$) is not controlled by the array. It includes energy of neutrinos, energy transferred to nuclei in various reactions and ionization losses of muons and hadrons in the atmosphere. In~\cite{Yakutsk:VI-ECRS(1978), Glushkov:PhD(1982)} this value was obtained from earlier calculations and is roughly consistent with predictions obtained with CORSIKA.

\section{Discussion}

Summary values of all constituents are given in the rightmost column of the Table~\ref{t:2}. The value $\E = 1.173 \times 10^{18}$~eV in the ``experiment'' column exceeds the mean value $\left<\E\right> = 0.997 \times 10^{18}$~eV obtained in simulation by factor $\simeq 1.177$. This difference is a result of overestimation of the scaling factor $k$, occurred in~\cite{Yakutsk:VI-ECRS(1978), Glushkov:PhD(1982)}, where it was determined as $k = 3.7 \times 10^{4}$~eV/photon\,eV$^{-1}$, while simulation with CORSIKA gave $\left<k\right> = 3.157 \times 10^{4}$~eV/photon\,eV$^{-1}$.

\begin{figure}
    \centering
    \includegraphics[width=0.85\textwidth, clip]{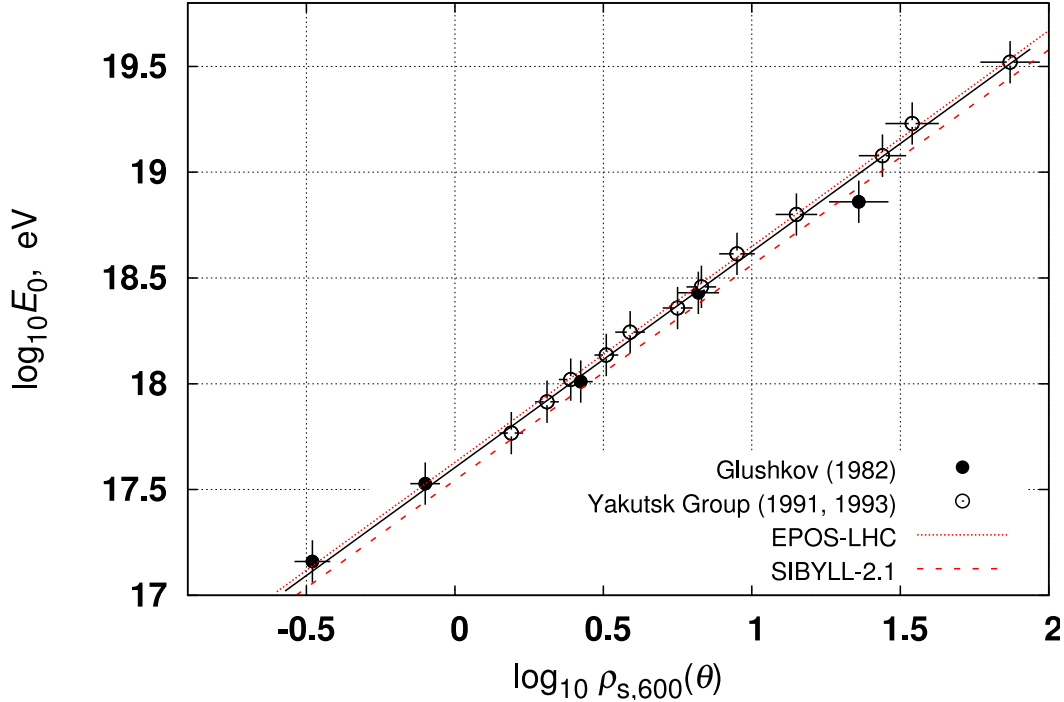}
    \caption{Energy $\E$ reconstructed from the parameter $\rhos(\theta)$ in showers with $\left<\cos{\theta}\right>=0.95$. Symbols represent the data given in~\cite{Glushkov:PhD(1982)} and~\cite{Yakutsk:Bull.Acad.Sci.USSR(1991), Yakutsk:TokyoWorkshop(1993)} reprocessed with the new calorimetric estimation. Solid line represents the best fit to all data.}
    \label{fig:s600_E0}
\end{figure}

The new estimation of primary energy obtained with the use of calorimetric method described above is given in the bottom row of the Table~\ref{t:2}. The value $\E = 1.019 \times 10^{18}$~eV was determined with corrected values $E_i = \left<k\right>\cdot F$, $\left<E_{1\mu}\right> = 10.3$~GeV and $\Delta E$. It is shown on Fig.\ref{fig:s600_E0} together with other data from~\cite{Glushkov:PhD(1982)} with black circles. White circles represent the data from~\cite{Yakutsk:Bull.Acad.Sci.USSR(1991), Yakutsk:TokyoWorkshop(1993)} reprocessed with the revised values of $F$ and $E\ionize$ with the account of the adjusted atmosphere transparency and with introduction of a new scaling factor $k$ (see Fig.\ref{fig:scaling}). Solid line represents the dependency:
\begin{equation}
    \E = (3.60 \pm 0.3) \times 10^{17} \cdot (\rhos(0\degr))^{1.02 \pm 0.02}\text{,}
    \label{eq:EnergyCalorimetric}
\end{equation}
which describes all the experimental data when $\rhos(18.2\degr)$ is converted to vertical with the use of (\ref{eq:rho_vert1982}) with $\lambda_{\rho} = 415$~\depthunit. Dotted and dashed lines reflect the relations (\ref{eq:EnergySIBYLL}) and (\ref{eq:EnergyEPOS}) which signify limits of the interval containing predictions of all the abovementioned models. The closest to experiment are \QGSII{} and \EPOS{} though one cannot exclude the credibility of two others.

\begin{figure}
    \centering
    \includegraphics[width=0.85\textwidth, clip]{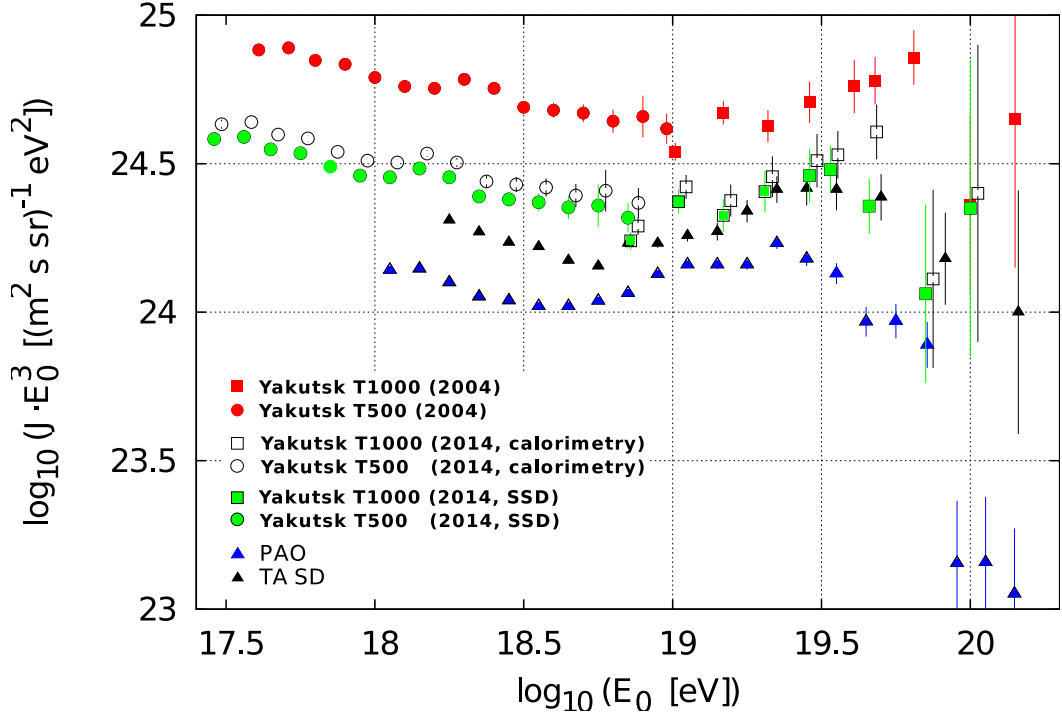}
    \caption{Differential energy spectrum of CR according to the data from different experiments.}
    \label{fig:CR_Spectrum}
\end{figure}

On Fig.\ref{fig:CR_Spectrum} the CR energy spectra are shown measured by modern giant EAS arrays. Circles and squares (showers selected by master triangles with $500$~m and $1000$~m sides correspondingly) represent the data of the Yakutsk experiment~\cite{Yakutsk:Nucl.Phys.B.Proc.Suppl.(2004)}. Energy $\E$, estimated with the use of expressions (\ref{eq:EnergyEst1993}) and (\ref{eq:FreePath1993}), are shown with red symbols; open symbols represent the same data with energy estimated according to (\ref{eq:EnergyCalorimetric}); green symbols~--- according to (\ref{eq:EnergySSD}). Black triangles represent TA SD data~\cite{TA(2011)}, blue triangles~--- PAO~\cite{PAO(2011)}.

\section{Conclusion}

Application of the CORSIKA code to the Yakutsk EAS array data provided an opportunity to critically examine the experiment's energy calibration which for a long time has been a subject of debates and controversy among our colleagues from other world EAS arrays. This became possible thanks to the availability of modern EAS development models to a wide range of researchers. With these models we have managed to calculate the responses of scintillation detectors and obtain a set of probable estimations for primary energy~(\ref{eq:EnergyQGS01}-\ref{eq:EnergyEPOS}). Calculations have revealed, that in relations (\ref{eq:EnergyEst1982}) and (\ref{eq:EnergyEst1993}) the energy dissipated in the atmosphere in the form of electromagnetic component, was overestimated by $(12-17)$\% depending on the shower maximum $\xmax$ (see Fig.\ref{fig:scaling}). This was made worse in (\ref{eq:EnergyEst1993}) due to underestimation of the atmosphere transparency by $\simeq 17$\%. The new calorimetry~(\ref{eq:EnergyCalorimetric}) has lead to a lower estimated value of $\E$ in comparison with~(\ref{eq:EnergyEst1993}) by factor $\simeq 1.33$ and in decreased intensity of the CR energy spectrum measured on the Yakutsk EAS array (see Fig.\ref{fig:CR_Spectrum}).

Independent techniques of $\E$ estimation from SSD LDFs (\ref{eq:EnergySSD}) and with the use of calorimetric method (\ref{eq:EnergyCalorimetric}) gave close results, which agree with simulations within $(10-15)$\%. At $\E \ge 8 \times 10^{18}$~eV they do not contradict the TA data~\cite{TA(2011)} and consistently point at the steepening of the primary CR spectrum in the region of extreme energies ($\E \ge 3 \times 10^{19}$~eV). This steepening does not contradict to GZK cutoff but, probably, has a different astrophysical reason. As for the difference in spectral intensities at $\E \le 8 \times 10^{18}$~eV, it could have other reasons. Probably, it is the effect of systematical errors in primary energy reconstruction techniques adopted by different experiments. But one cannot exclude that the said difference is caused by geographical locations of arrays observing different regions of the sky. In~\cite{Glushkov&Pravdin:JETPLett(2008)} such a correlation had been noticed. Our current plan is to continue the elaboration of $\E$ estimation at the Yakutsk array with a more detailed analysis of the Cherenkov light data.

\acknowledgements

This work is financially supported by Russian Academy of Science within the program ``Fundamental properties of matter and astrophysics'', by RFBR grant 13-02-12036~ofi-m-2013 and by grant of President of the Republic Sakha (Yakutia) to young scientists, specialists and students to support research.

\bibliographystyle{aipnum4-1}
\bibliography{energy-estimation-2014}

\end{document}